\documentclass{JINST}
\usepackage{graphicx}
\usepackage{epsfig}
\usepackage{amssymb}
\usepackage{topcapt}
\usepackage{bm}
\usepackage{latexsym}
\usepackage{subfigure}
\usepackage{lscape}
\usepackage{footnote}
\def \Kr85 {$^{85}$Kr }

\newcommand{\footnoteremember}[2]{
\footnote{#2}
\newcounter{#1}
\setcounter{#1}{\value{footnote}}
}
\newcommand{\footnoterecall}[1]{
\footnotemark[\value{#1}]
}

\title{In situ measurements of Krypton in Xenon gas with a quadrupole
  mass spectrometer following a cold-trap at a temporarily reduced
  pumping speed}
\author{Ethan Brown\thanks{Corresponding author.}, Stephan Rosendahl,
  Christian Huhmann, Christian Weinheimer, Hans Kettling\\
Institut f\"ur Kernphysik, University of M\"unster,
Wilhelm-Klemm-Str. 9, D-48149 M\"unster, Germany \\
Email: \email{ethanbrown@uni-muenster.de}}


\abstract{A new method for measuring trace amounts of krypton in xenon
  using a cold trap with a residual gas analyzer has been developed,
  which achieves an increased sensitivity by temporarily reducing the
  pumping speed 
  while expending a minimal amount of xenon.  By partially closing a
  custom built butterfly valve between the measurement chamber and the
  turbomolecular pump, a sensitivity of $40$ ppt has been reached. 
  This method has been tested on an ultra-pure gas sample from Air
  Liquide with an unknown intrinsic krypton concentration, yielding a
  krypton concentration of $330 \pm 200$ ppt.}

\begin{document}

\section{Introduction}

Xenon detectors are used for a large number of low background particle
physics experiments, including direct dark matter detection,
neutrinoless double beta decay, and solar neutrino searches
\cite{XENON100} - \cite{DMReview}.  Such
experiments, which look for extremely rare interactions, require an
ultra low intrinsic background.  
Xenon detectors are currently among the most sensitive detectors for
these type of experiments, and have pushed the technological and
scientific limits in various fields \cite{XENON100PRL} \cite{EXO2nu}
for several reasons.  Due to the high Z value, xenon
detectors shield external gamma backgrounds at the central region of
the detector.  Xenon also has a low intrinsic radioactivity,
as it has no long lived radioactive isotopes, except for $^{136}$Xe
\cite{EXO2nu} \cite{Kamland2nu},
which is the subject of double beta decay searches, and hence has a
low enough rate to be ignored as a background in other experiments at
current sensitivities.
Additionally, xenon detectors can be scaled to large volumes with
relative ease, allowing for incredible sensitivity for these rare
event searches on a short time scale.

However, to achieve a sufficiently low background for these rare event
searches, at the level of $10^{-2}$ dru
\footnoteremember{ft:dru}{differential rate unit: 1 dru = 1 event /
  keV / kg / day}, internal impurities must be 
reduced substantially relative to what is available in commercial
xenon.  One such impurity that has an important impact on xenon
detectors is $^{85}$Kr, which has a beta decay with an endpoint energy of
687 keV. \Kr85 is created in nuclear fuel reprocessing and was formerly
created in nuclear weapons testing, but with a relatively long half
life of 10.8 years, both sources contribute to the present
concentration in the atmosphere at a level of
$10^{-11}$ \Kr85 in Kr.  \Kr85 must be removed from commercial xenon
to achieve the low 
internal background necessary for modern astroparticle physics
experiments.  Comercially available xenon can be purchased with an
intrinsic contamination of less than 10 ppb
\footnoteremember{ft:ppb}{1 ppb = $10^{-9}$ mol/mol} Kr in Xe, but the
low background requirements 
dictate that this fraction must be further reduced to levels below 1
ppt \footnoteremember{ft:ppt}{1 ppt = $10^{-12}$ mol/mol}.  

\Kr85 is commonly removed from xenon by reducing the total amount of
Kr in the gas via cryogenic distillation \cite{Abe2009290}.  It is of
key importance to be able to measure the amount of krypton in xenon,
both for the input gas at the level of 10 ppb, as well as down to the
best purity levels currently achievable at the sub-ppt level.  While
several methods exist to make this measurement, we present here a
method that can be performed with common equipment used in gas
laboratories with nearly zero xenon consumption.  Moreover, this is a
fast, online method that can be performed in situ, as opposed to
offline methods that require up to weeks to obtain a result.

\section{Enhanced Gas Content Measurements using a Residual Gas Analyzer with a Liquid Nitrogen Cold Trap}

Mass spectrometry devices are capable of measuring the mass
composition of gas samples, but due to the limitations of the dynamic
range of such devices, they cannot be used in standard operation to
measure trace impurity concentrations at the ppt level.  One technique
to enhance the sensitivity of this measurement is to use a liquid
nitrogen cold trap to reduce concentration of the dominant gas species
by orders of magnitude, thereby allowing a quantitative measurement of
the trace impurities without saturating the mass spectrometer.  This
method has already been developed to measure various impurities in
xenon gas \cite{Leonard2010678} - \cite{Dobi201240}, and has been
investigated further for the particular case of measuring the
concentration of krypton in xenon at the sub ppt level
\cite{Dobi20111}.  

\begin{figure}[!h]
  \centering
  \includegraphics[width=.9\linewidth]{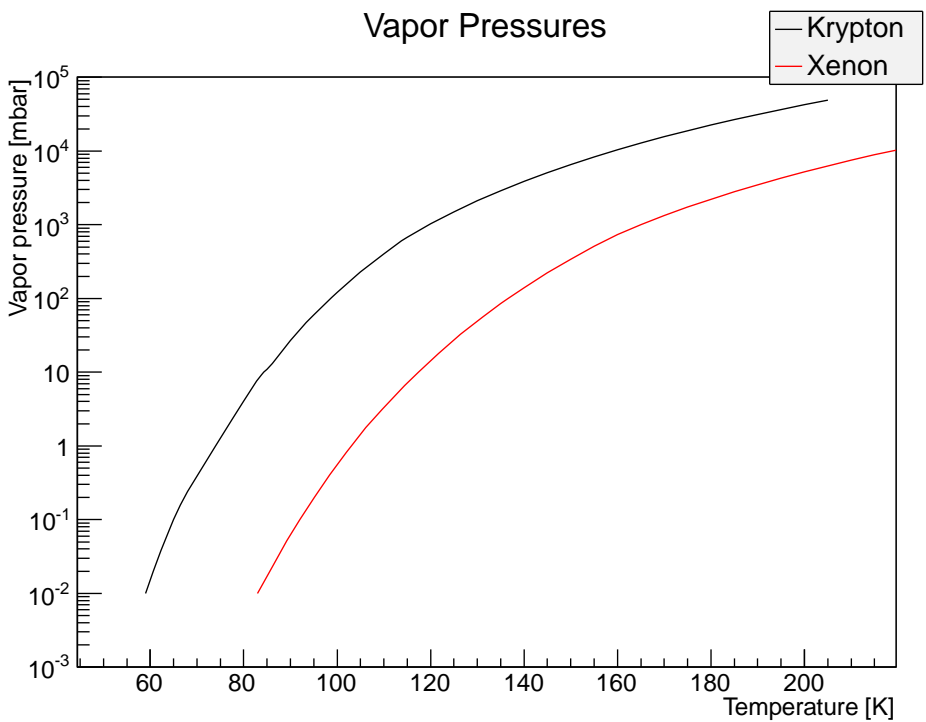}
  \caption{Vapor pressures of xenon and krypton
    \protect\cite{vaporpressure}.  While no data are available for
    xenon at 77 K, extrapolation of the xenon curve yields a value of
    $2.5 \times 10^{-3}$ mbar, which is consistent with the 3 order of
    magnitude difference between xenon and krypton at slightly higher
    temperatures.}
  \label{fig:vaporpressure}
\end{figure}

The principle behind the sensitivity enhancement of this measurement
lies in the difference of vapor pressures of xenon and the impurities
of interest at 77 K, in this case krypton.  At this temperature, xenon
has a vapor pressure of $2.5 \times 10^{-3}$ mbar, while that of
krypton is orders of magnitude higher at $2.0$ mbar
\cite{vaporpressure}, as shown in figure
\ref{fig:vaporpressure}.  Independent of the incoming
flow rate, the cold trap reduces the xenon pressure to the vapor
pressure, allowing the krypton to pass through more or less
unattached \footnoteremember{ft:vap}{The concept of vapor pressure
  describes the equilibrium between the gaseous phase of an atom or
  molecule with the solid or liquid phase, It is governed by the
  atom-atom or molecule-molecule binding energies in the liquid or
  solid phase but it neglects surface-atom or surface-molecule binding
  effects, which depend very much on the surfaces of the
  recipient. Only if one or more monolayers of the atom or molecule is
  covering the walls the atom-atom or molecule-molecule interaction
  takes over and vapor pressure is a good concept. Therefore, we
  expect deviations from this simple concept of the vapor pressure at
  low amounts of atoms or molecules.}.  This allows enough krypton to
be introduced to the mass spectrometer to measure the trace
concentration while maintaining the low pressure necessary to operate
the device.

In this work, modifications are introduced that address two important
issues.  First, the amount of gas used for the measurement is
drastically reduced, allowing for many measurements without expending
large amounts of xenon.  This is of key importance for applying this
method for long-term quality assurance purposes on large quantities of
xenon, since xenon is quite expensive.  This is realized by reducing
the flow rate of xenon into the system and also by making measurements
on a relatively short time scale.  Using this approach it is possible
to measure the krypton concentration in a very nearly non-destructive
way. 

The second issue addressed here is an additional sensitivity
enhancement by temporarily reducing the pumping speed in the
measurement chamber. Mass spectrometry devices must typically be
operated at a pressure below $~10^{-5}$ mbar, and are therefore
usually installed very near a turbomolecular pump (TMP), ensuring that the
vacuum stays well below the required limit.  By partially closing a
valve in front of the turbo pump for a short time, the pressure rises,
thus enhancing the sensitivity to all gas species.  In order to
maintain safe operation of the mass spectrometer, the scan range is
set only to the trace masses of interest, so that the increased
pressure of the bulk gas components does not saturate the device.  

Finally, in order to make quantitative analyses of the Kr/Xe fraction
in gas samples in the ppb range and below, a calibration method has
been developed where krypton can be artifically doped in a xenon
sample over a large range of concentrations.  To achieve this, a
volume division method has been established, which allows doping at
the ppb level and facilitates a quantitative measurement of krypton in
xenon to levels below 1 ppb.

\section{Experimental Setup}

\begin{figure}[!h]
  \centering
  \includegraphics[width=.9\linewidth]{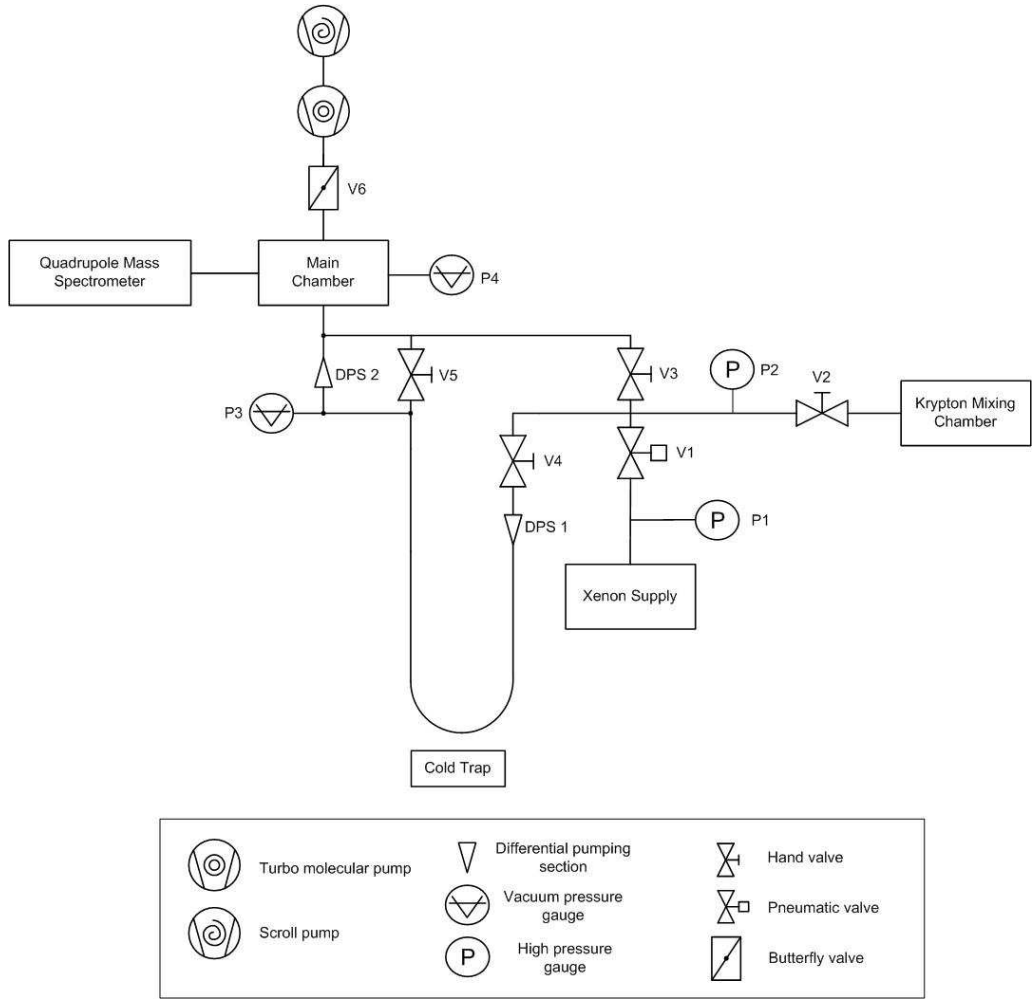}
  \caption{Diagram of experimental setup.}
  \label{fig:setup}
\end{figure}

A diagram of the experimental apparatus used for our measurements is
shown in figure \ref{fig:setup}.  The gas sample is introduced by
opening valve V1, after which the gas passes a low conductance
differential pumping section before entering a liquid nitrogen cold
trap.  There is an additional low conductance differential pumping
section between the cold trap and the main chamber where the
measurement is performed. The main chamber houses a Transpector II
quadrupole mass spectrometer \cite{Inficon}.  A custom made butterfly
valve is mounted between the main chamber and the TMP
used to evacuate the system. 

\begin{figure}[!h]
  \centering
  \includegraphics[width=.9\linewidth]{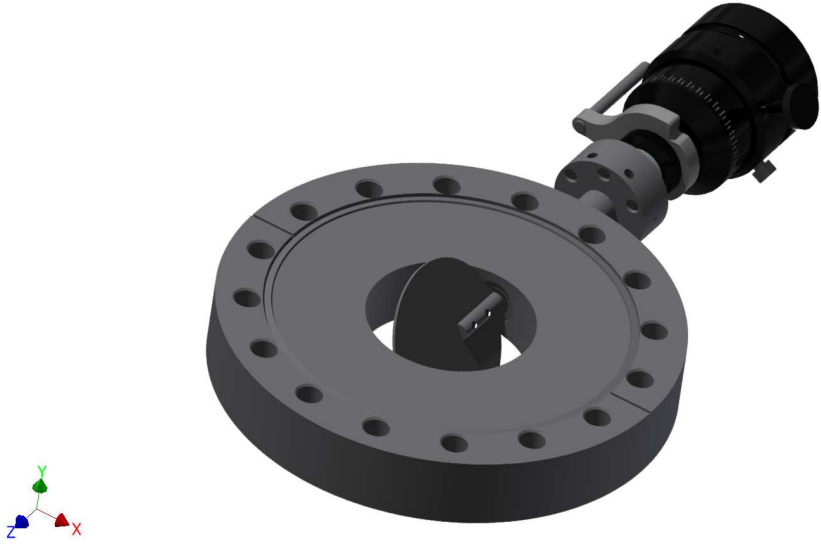}
  \caption{Custom built butterfly valve, which partially closes the
    opening between the main chamber and the TMP, reducing the
    effective pumping speed.  The position of the valve is controlled
    by a rotational feedthrough.}
  \label{fig:BV}
\end{figure}

The key difference between our setup and that described in reference 
\cite{Leonard2010678} is the presence of the custom made butterfly
valve located between the main chamber and the TMP, which allows to
reproducibly reduce the pumping speed, thereby enhancing the
sensitivity to trace gases.  As shown in figure \ref{fig:BV}, the
butterfly valve consists of a cylindrical plate 1 mm thick with a 54.6
mm diameter mounted on a rotational feedthrough.  The plate can be
turned to partially block the opening in the CF100 flange in which it
is mounted, whose inner diameter is 55 mm.  The position of the
butterfly valve can be controlled with better than 1$^\circ$ precision.  

In order to optimize the sensitivity enhancement, the mass
spectrometer can be operated only in the mass range of the trace gas
of interest.  In the case of krypton in xenon, where the dominant gas
component is xenon, limiting the range from 76 to 90 atomic mass units
(amu) allows operation of the mass spectrometer at pressures above the
specification of $10^{-5}$ mbar with a high sensitivity to krypton
without damaging the ion detection electronics since the partial pressure remains sufficiently low.


An understanding of the gas flow through the system is outlined here
by treating only the xenon flow through the system.  While there is a
mass dependence of the gas flow, an understanding of the xenon flow
through the system is sufficient to conduct the measurements.  The
different behavior of krypton gas in the system is accounted for by
using calibrations with krypton gas at many different concentrations.
This allows the behavior of krypton relative to xenon to be absorbed
into the fit parameters introduced in the analysis. 

The gas flow through the system is naturally divided into two sections at
the cold trap, since most of the gas freezes at this point.
Before the cold trap, the flow dynamics are dominated by the limited
conductivity through the first differential pumping section DPS1 while
after the cold trap the conductivity of the second differential
pumping section and butterfly valve are most important.

\begin{figure}[!h]
  \centering
  \includegraphics[width=.9\linewidth]{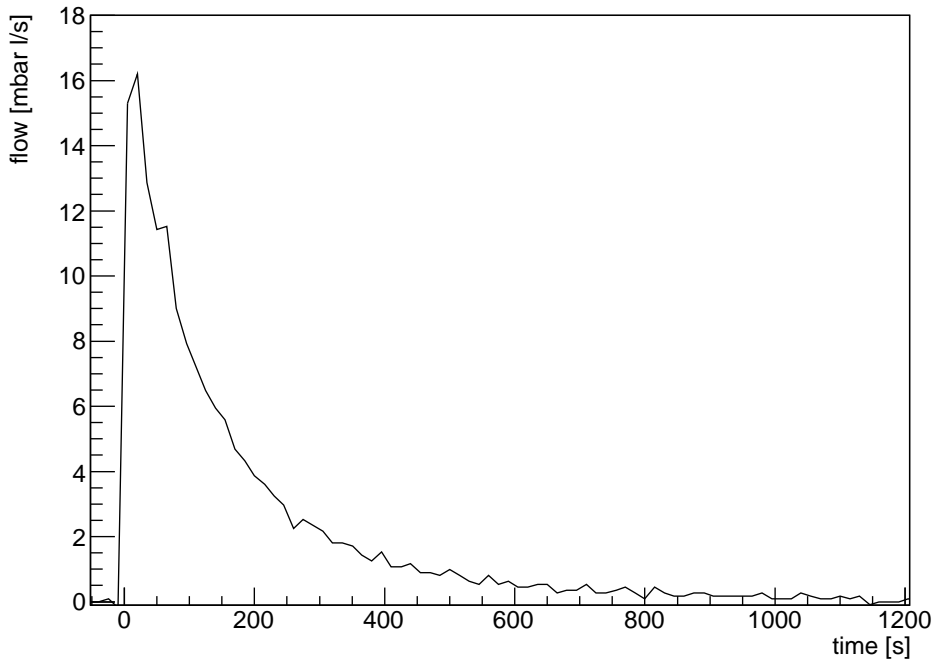}
  \caption{Flow evolution at the gas inlet for an example measurement
    as measured by the time rate of change of the inlet pressure.}
  \label{fig:flowin}
\end{figure}

\begin{figure}[!h]
  \centering
  \includegraphics[width=.9\linewidth]{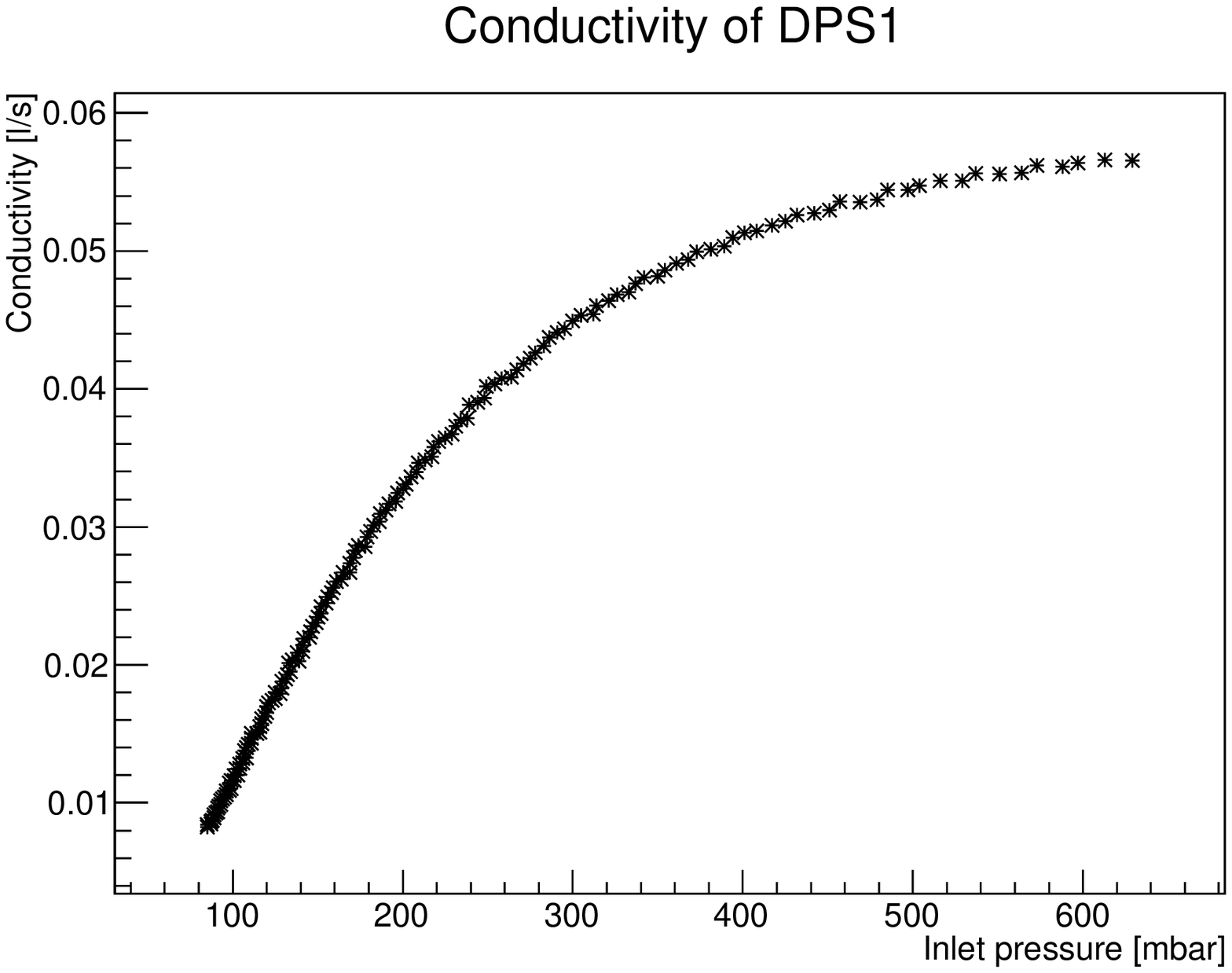}
  \caption{Conductivity $C_1$ of DPS1 as a function of inlet
    pressure as measured by equation
    (\protect\ref{eqn:qinmeasure}). The linear dependence of 
    the conductivity at low inlet pressure is typical of laminar
    non-turbulant flow.  The non-linear behavior at higher pressure is
    due to the flow becoming turbulent.}
  \label{fig:Cin}
\end{figure}

In the region before the cold trap, the flow can be measured simply by
the time rate of change of the inlet pressure $p_{in}$.  For a fixed
volume of gas $V$, the flow  $q_{in}$ is given by,
\begin{equation}
  q_{in} = -V\frac{\mathrm{d}p_{in}}{\mathrm{d}t}.
  \label{eqn:qin}
\end{equation}
Figure \ref{fig:flowin} shows the flow rate of a sample
measurement.  Additionally, the flow is related to the conductivity of
the differential pumping section $C_1$ by,
\begin{equation}
  q_{in} = C_1 \Delta p \approx C_1 p_{in},
  \label{eqn:qinmeasure}
\end{equation}
since the pressure after the leak valve is much smaller than
$p_{in}$.  Alternatively, the conductivity can be calculated in the
laminar flow regime by,
\begin{equation}
  C_1 = \frac{\pi~d^4}{256~\eta~l} p_{in},
  \label{eqn:c1}
\end{equation}
where $d = 0.13$ mm is the inner diameter and $l = 100$ mm is the
length of the differential pumping section, and $\eta = 2.3 \times
10^{-5}$ Pa s is the viscosity of xenon at 293 K.
The conductivity of DPS1 is shown in figure \ref{fig:Cin} as measured
by equation (\ref{eqn:qinmeasure}).  At low inlet pressures, a
linear behavior of the conductivity is observed, typical of laminar
flow.  While the precise value of the conductivity 
is not of crucial importance, a conductivity in this range is
necessary to optimize the sensitivity to krypton and the consumption
of xenon. 

The Reynolds number, which is used to determine when a flow
becomes turbulent, is given by
\begin{equation}
  Re = \frac{4 \rho q}{\pi \eta p_{in} d},
  \label{eqn:Reynolds}
\end{equation}
where $q$ is the flow rate.  A Reynolds number below 2500 indicates
the flow is laminar, and a Reynolds number above 4000 indicates it is
turbulent, while intermediate values correspond to a transitional
regime.  At low inlet pressures around $p_{in} =
100$ mbar, the Reynolds number is $Re = 250$, consistent with laminar flow.
At $p_{in} = 300$ mbar, where the conductivity begins to show
non-linearity, the Renolds number is $Re = 3000$.  At a higher inlet
pressure of $p_{in} = 600$ mbar, a value of $Re = 7900$ is consistent
with turbulent flow.

At the cold trap, where the bulk of the gas freezes, the partial
pressure of each gas does not exceed the vapor pressure at liquid
nitrogen temperature.  We consider here the cases of xenon and
krypton.  

Xenon has a vapor pressure of $2.5 \times 10^{-3}$ mbar at 77 K while
that of krypton is  $2.6$ mbar, which allows the krypton to pass
through the cold trap unattached, since the partial pressure of
krypton is always well below this value.  As mentioned in footnote
\footnoterecall{ft:vap}, this is not fully true, since some krypton
can attach to the wall of the cold trap by atom-surface interactions.
Indeed, after warming up the cold trap, we see some krypton coming out
of the trap.  Even for a high krypton
concentration, say from a doping measurement with 10000 ppb Kr/Xe, the
krypton partial pressure in the inlet volume is around $10^{-2}$ mbar,
well below the vapor pressure.  Since the pressure is substanially
reduced between the inlet and the cold trap, the partial pressure of
krypton in
the cold trap is always substantially below this.  In practice, the
pressure in the cold trap is slightly higher than the xenon vapor
pressure, and is measured to be $2.5 \times 10^{-2}$ mbar, but since
the pressure in the cold trap was monitored a full understanding of
the gas dynamics is still possible.  In principle, a cold trap with a
larger surface area may fully freeze the xenon, but this pressure
reduction was adequate to perform measurements.  

The total gas
pressure is then further reduced in the measurement chamber by the use
of a second low conductance differential pumping section, DPS2.  Since
at these low pressures the gas flow is in the molecular flow regime,
it is possible to calculate the conductivity $C_2$ through DPS2, which
is given by
\begin{equation}
  C_{2} = \frac{\pi}{16}\cdot \bar{c} \cdot d^2 \frac{14+4\frac{l}{d}}
  {14+18\frac{l}{d}+3(\frac{l}{d})^2}
  \label{eqn:conductivity}
\end{equation}
where $l = 5$ mm is the length of the tube and $d = 1$ mm is the
diameter, and 
\begin{equation}
  \bar{c} = \sqrt{\frac{8RT}{\pi M}}
  \label{eqn:cbar}
\end{equation}
is the mean particle speed.  We use a value of $\bar{c} = 217$ m/s for
xenon at $T = 293$ K, which corresponds to a conductivity of $C_{2} = 8
\times 10^{-3}$ l/s.  

Note that even with the use of the cold trap and the much higher vapor
pressure of krypton relative to xenon, the dominant component of the
gas is still xenon.  This is due to the fact that the krypton
concentration is so low.

The xenon ice in
the cold trap provides a constant pressure source of gas to the
differetial pumping section, which enters the main chamber and is
pumped away by the TMP.  The flow into the main
chamber $q_{MC}$ is given by equation (\ref{eqn:qinmeasure}), with $C_{1}$
replaced by $C_{2}$ and $p_{in}$ replaced by the pressure in the cold
trap $p_{CT} = 2.5 \times 10^{-2}$ mbar, yielding a flow of $q_{MC} = 2 \times 10^{-4}$ mbar l/s. 

The flow out of the main chamber $q_{out}$ is determined by the
pressure in the main chamber $p_{MC}$ and the effective pumping speed
of the TMP $S_{eff}$, 
\begin{equation}
  q_{out} = p_{MC} \cdot S_{eff},
  \label{eqn:qout}
\end{equation}
The effective pumping speed is reduced from the full pumping speed $S$
due to the presence of the butterfly valve between the TMP and the
main chamber, which has a conductivity $C_B$, and is related by,
\begin{equation}
  \frac{1}{S_{eff}} = \frac{1}{C_B} + \frac{1}{S}.
  \label{eqn:Seff}
\end{equation}

An estimation of conductivity of the butterfly valve in the fully open
position can be found by treating it as a simple aperture, and is
given by
\begin{equation}
  C_B = \frac{\pi d^2}{16} \bar{c},
  \label{eqn:CB}
\end{equation}
where $d$ is the diameter of the aperture and $\bar{c}$ is the
mean particle speed given in equation (\ref{eqn:cbar}).  The aperture
of the butterfly valve has a diameter of $d = 55$ mm, which
corresponds to a conductivity of $C_B = 130$ l/s. With a pumping speed
of $S = 300$ l/s, this corresponds to an effective pumping speed of
$S_{eff} = 90$ l/s.  The pressure in the main chamber during a
measurment with the butterfly valve fully open is around $p_{MC} = 1.5
\times 10^{-6}$ mbar, yielding a flow out of the chamber of $q_{out} =
1.4 \times 10^{-5}$ mbar l/s, in fair agreement with the flow into the
chamber.

The final dynamical consideration of the system is the gas flow out of
the turbo pump when the butterfly valve is partially closed.  With the
butterfly valve set to 14$^\circ$ from fully closed, the setting used
for the measurements, the pressure in the main chamber was $p_{MC} =
5 \times 10^{-5}$ mbar.  With this, the effective pumping speed for xenon
is reduced by a factor of 30, to $\tilde{S}_{eff} = 3$ l/s.

Calibrations were performed using xenon samples artifically doped with
krypton.  The doping proceeded by use of volume division.  A sample of
krypton was placed in a small volume at 1.0 bar, and then expanded
into a second volume which was larger by a factor of 5.6.  The large
volume was then evacuated with a TMP and the process
was repeated.  The pressure in the small volume was monitored with a
Baratron MKS type 121A pressure sensor, which is accurate to $0.5\%$
in the range of 1 mbar to 10 bar.  The pressure sensor eliminates the
need for a precise calibration of the volume sizes, but such a
calibration was performed as a cross check and was found to be
consistent.  Finally, the krypton at a reduced pressure is mixed with
xenon in an equally sized volume at 2 bar to allow for a precise
doping to levels of $10^{-3}$.  To achieve doping levels at lower
concentrations, the volume division process can be repeated on a gas
sample at a doping level around $10^{-3}$, to reach lower doping
concentrations. 

\section{Measurements}

To perform a measurement, a gas sample with known doping is prepared.
The cold trap is prepared with the liquid nitrogen filled to roughly
the same level around the trap for 10 minutes before beginning the
measurement.  Repeated measurements show that the precise level of the
nitrogen is not important.  The RGA is set to scan only the trace
krypton isotopes, a range from 76 to 90 amu, which avoids potential
saturation effects in the RGA at xenon mass units.  

The gas sample is prepared and its pressure is monitored by the
pressure sensor P2.  Just before beginning a measurement, the pumping
speed of the TMP is reduced by partially closing the butterfly valve
to 14$^\circ$ from fully closed.  The gas sample is then introduced
into the system by opening valve V4, and the mass spectrum is recorded
in the pre-defined range.  The gas is fed into the system for several
minutes, and the hand valve V4 is then closed.  The pressure in the
main chamber is always monitored to ensure that the total pressure
stays in a safe range for operation of the RGA.   

The RGA records the current at steps of 0.2 amu.  The current
for a given mass number is found by integrating over three steps with
0.2 amu difference centered around the mass of interest.  This was
chosen to optimize the signal to noise ratio.

\begin{figure}[!h]
  \centering
  \includegraphics[width=.9\linewidth]{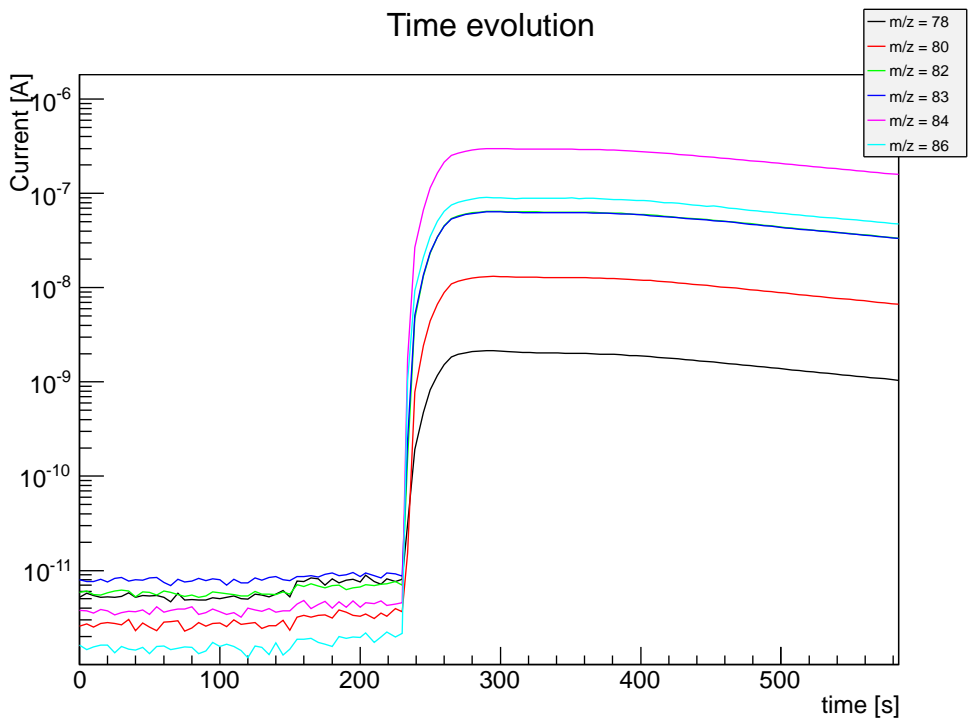}
  \caption{Example measurement showing the time evolution of krypton
    isotopes}
  \label{fig:measurement}
\end{figure}

An example measurement is shown in figure \ref{fig:measurement}.  In
this case, a gas sample with a krypton doping of 53000 ppb was used.
The gas sample is introduced at $t = 200$ s, and the krypton signal
appears in the RGA after a short delay.  This delay is dependent on
the krypton concentration and is longer at lower concentrations.  The
delay is likely due to surface effects, where a finite amount of
krypton attaches to the surface of the cold trap before equilibrium is
reached, but after this short time the remainder of the krypton passes
through unattached \footnoterecall{ft:vap}.  The figure
shows the time behavior of several krypton isotopes, which are treated
in detail in section \ref{sec:Analysis}.  The signal peaks shortly
after the sample is introduced, then decreases as the input pressure
drops, and hence the flow rate is reduced.  

\begin{figure}[!h]
  \centering
  \includegraphics[width=.9\linewidth]{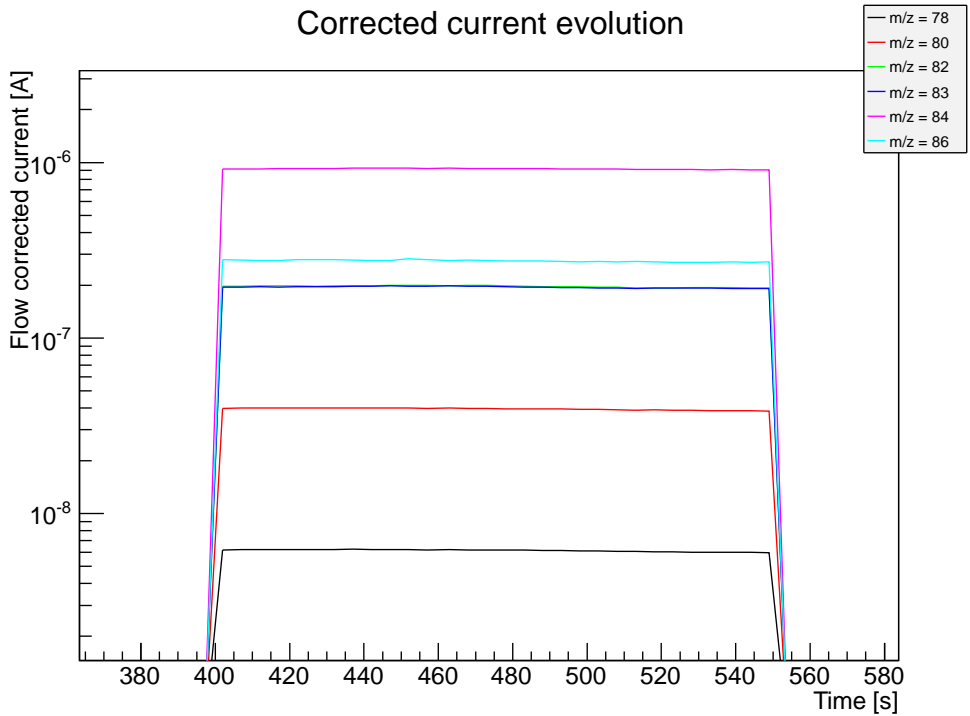}
  \caption{Example measurement with normalization by flow rate.  Note
    that the relative currents for the different isotopes follow the
    expectations presented in table \protect\ref{tab:abundance} below.}
  \label{fig:PNorm}
\end{figure}

Since the input pressure and gas flow are not constant over the
measurement, a correction must be made in order to perform a proper
quantitative analysis of the krypton concentration.  The current is
normalized by the input flow between $t = 400$ s and $t = 550$ s by
applying a correction
\begin{equation}
  I_c(t) = I(t) \times \frac{q(t)}{q_0},
  \label{eqn:IC}
\end{equation}
where $q_0 = 10$ mbar l/s is used as a refernece flow rate.  The
resulting flow corrected current is shown in figure \ref{fig:PNorm}.
The corrected current is nearly 
flat after normalization, thus confirming that this yields the correct
time-dependent correction.  For analyzing the measurements, this
region is averaged.  

One important aspect of this measurement technique is that it consumes
only a small amount of xenon.  Since xenon is expensive, it is useful
to be able to perform a quantitative gas analysis repeatedly without
expending a large amount of gas.  

For calibration measurements with high doping concentrations and for
undoped measurements, the amount of xenon gas required is minimal.
Only that necessary to fill the inlet volume plus feed lines to a
supply bottle is required.  In practice, volumes much smaller than one
liter can easily be constructed.  In our setup, the total volume was
around 30 ml, and xenon gas was introduced at 2 bar.  While some gas
remained in the inlet volume after the measurment, it was discarded
for practical reasons.  Thus, each measurement expended 0.06 standard
liters of xenon, but this can likely be reduced for future
measurements by designing a system where one does not need to exhaust
the remaining sample gas after the measurement and possibly by
recovering the xenon frozen in the cold trap.  For comparision, the
method presented by Dobi et al \cite{Dobi20111} consumes around 8
standard liters per measurement.  Thus, our measurement method is
essentially non-consuming compared to other similar measurement
methods. 

For the low concentration doping calibration measurements, more xenon
was necessary due to the extra step of mixing the dilute krypton with
xenon in the dilution procedure.  Here, the inlet volume must be
filled as many as three times, yielding a total of 0.18 standard 
liters for the measurment.  However, these calibration measurements do
not necessarily need to be repeated for measuring multiple gas
samples, so this is not a routine requirement for gas consumption.

\section{Analysis}
\label{sec:Analysis}

Several measurements were made with a sample of xenon purchased from
Air Liquide.  This high purity gas is specified to have a Kr/Xe
concentration below 10 ppb, but the absolute concentration is not
measured by the company.  We use these measurements to several ends,
as an illustration of this procedure for measuring trace amounts of
krypton in xenon, to infer the ultimate sensitivity of the
measurement, and also to determine the actual Kr/Xe concentration of
the Air Liquide gas.  This last point is important for the context of
using cryogenic distillation to further reduce the krypton
contamination, since the input concentration is of some importance.

In order to determine the unknown concentration in the xenon, several
measurements were made with the RGA cold trap setup with known
dopings at 61,000 ppb, 53,000 ppb, 28,000 ppb, 830 ppb, 700 ppb, 25
ppb, 10 ppb and 9 ppb, as well as two undoped samples.  

Within one measurement, additional information can be obtained by
treating each isotope separately, since different isotopes are present
at different concentrations for the same doping.  The krypton isotopes
considered in this analysis and their relative natural abundances are
listed in table \ref{tab:abundance}.
The factor of nearly 160 between  the abundance of $^{84}$Kr and that
of $^{78}$Kr allows the study of the response of the measurement
procedure to a large range of concentrations with a single measurement.  

\begin{table}[!h]
  \centering
  \begin{tabular}{|c|c|}
    \hline
    \textbf{Isotope} & \textbf{Abundance $f$} \\
    \hline
    $^{78}$Kr & 0.35\% \\ 
    $^{80}$Kr & 2.25\% \\ 
    $^{82}$Kr & 11.6\% \\ 
    $^{83}$Kr & 11.5\% \\ 
    $^{84}$Kr & 57.0\% \\ 
    $^{86}$Kr & 17.3\% \\
    \hline
  \end{tabular}
  \caption{Krypton isotopes and their relative natural
    abundance \protect{\cite{vaporpressure}}.}
  \label{tab:abundance}
\end{table}

To find the relationship between the corrected current $I$ and the
concentration $c$, all data points with and without doping that were
clearly above the background were fit with a linear function of the
form, 
\begin{equation}
  \centering
  I(c) = a \times f \times c = a \times f \times (d + c_0),
  \label{eqn:fitfunc}
\end{equation}
where $c = d + c_0$ is the total concentration, $d$ is the doping
concentration, $c_0$ is the intrinsic concentration, and $f$ is the
isotopic fraction.  The free fit parameters are the normalization
constant $a$ and the intrinsic concentration $c_0$.

\begin{figure}[!h]
  \centering
  \begin{minipage}[b]{.9\linewidth}
   \centering
   \subfigure[]{
     \includegraphics[width=.43\linewidth]{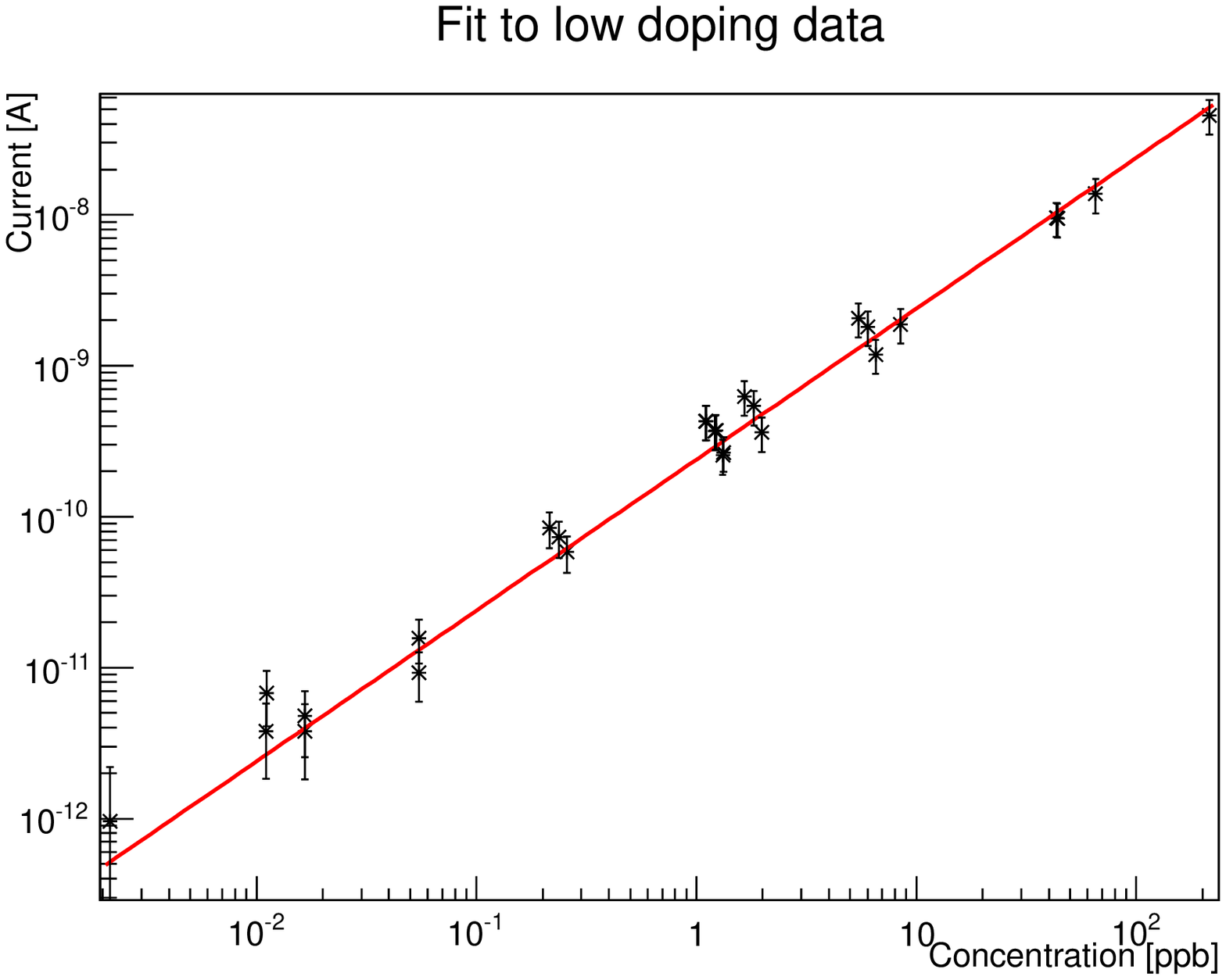}
     \label{subfig:lowDoping}
   }
   \subfigure[]{
     \includegraphics[width=.43\linewidth]{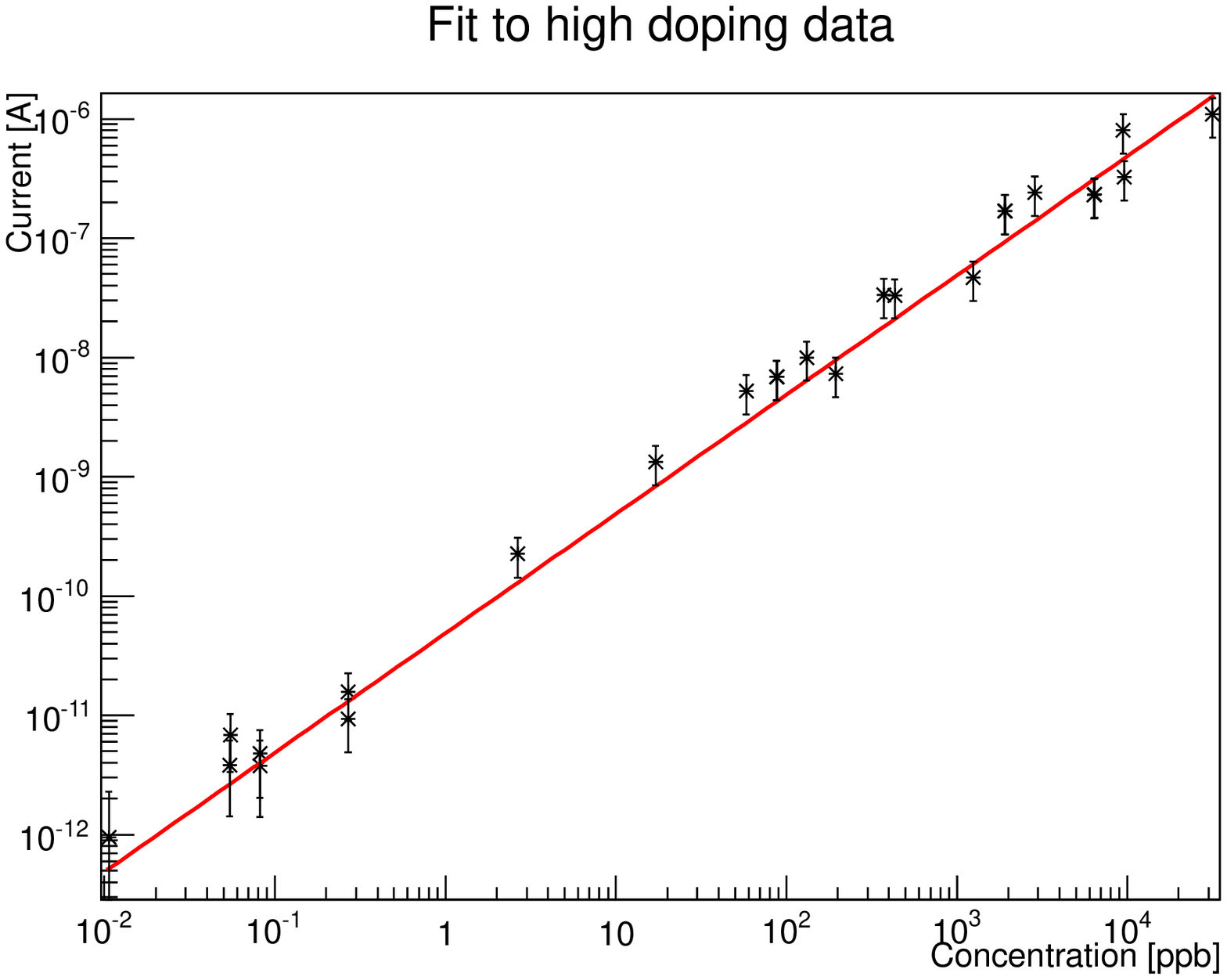}
     \label{subfig:highDoping}
    }
   \end{minipage}
   \begin{minipage}[b]{.9\linewidth}
     \centering
     \subfigure[]{
       \includegraphics[width=.9\linewidth]{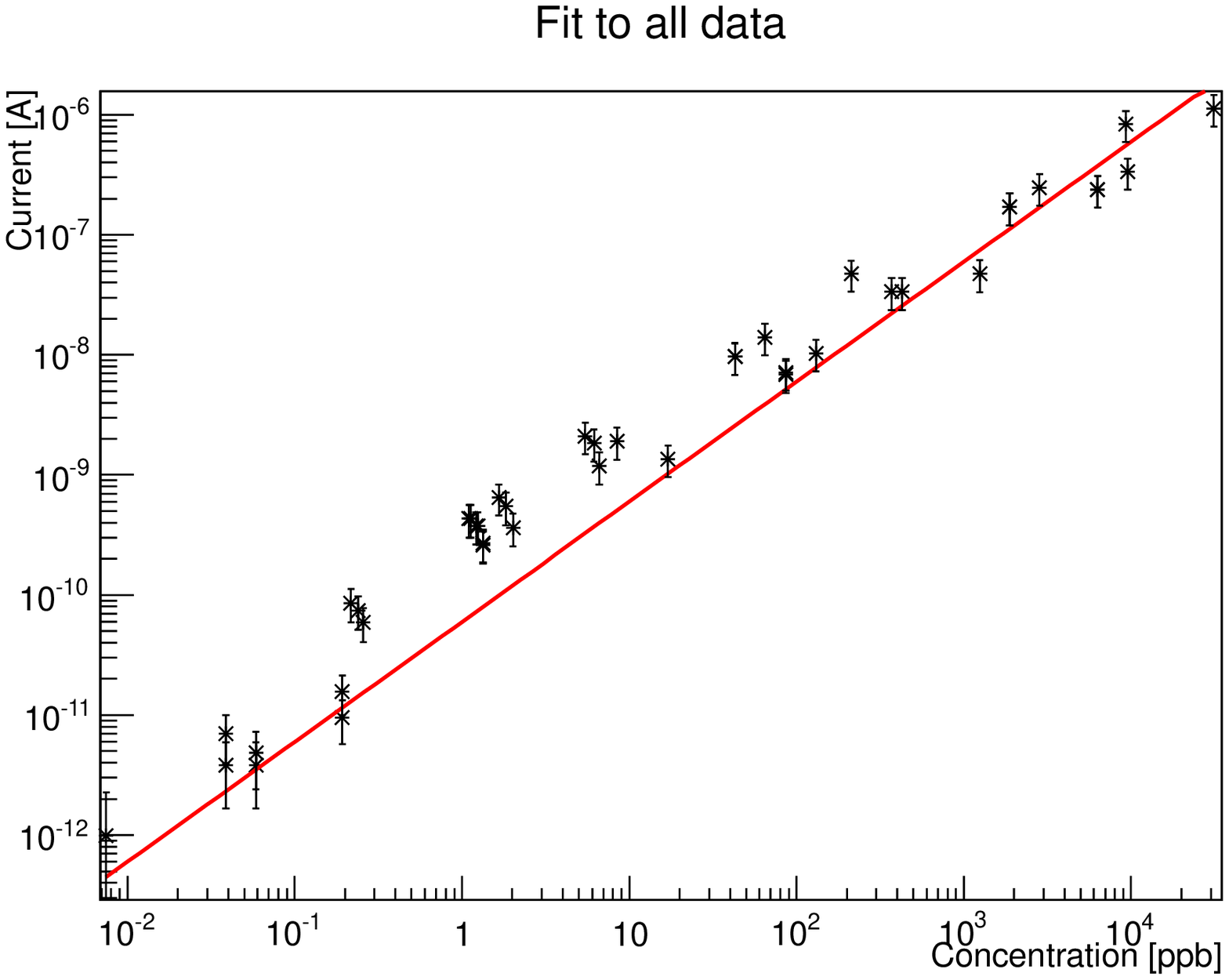}
       \label{subfig:fitall}
     }
   \end{minipage}
   \caption{Data from doped and undoped measurements with a linear
    fit with equation \protect\ref{eqn:fitfunc}.  Figure
    \protect\ref{subfig:lowDoping} shows only the low doping data,
    yielding fit parameters of $a = (240 \pm 10) \times 10^{-12}$
    A/ppb and $c_0 = 0.10 \pm 0.02$ ppb.  Figure
    \protect\ref{subfig:highDoping} shows only the high doping data,
    which gives fit parameters of $a =(49 \pm 4) \times 10^{-12} $
    A/ppb and $c_0 = 0.50 \pm 0.02$ ppb.  Figure \protect
    \ref{subfig:fitall} shows all data, with fit  paramters of $a =
    (59 \pm 3) \times 10^{-12}$ A/ppb and $c_0 = 0.34 \pm 0.21$
    ppb. The dominant error on $c_0$ comes from fitting to the
    different subsets of data.}
   \label{fig:FitAll}
 \end{figure}

Figure \ref{fig:FitAll} shows the data as well as linear fits to all
data and subsets thereof using equation \ref{eqn:fitfunc}.  
Of key interest is the fact that each measurement shows a linear
behavior, while there is a rather large discrepancy between different
measurements.  The data also lie in two distributions, one for the
high dopings and another for lower dopings.  This is likely due to a
systematic effect that is yet to be fully understood, but to show a
demonstration of the measurement method this is simply accounted for
in a systematic error, which is then the dominant error on the
measurement process.

Figure \ref{subfig:lowDoping} and \ref{subfig:highDoping} show the
fits to the low and high doping data separately.  The error is treated
as two terms, a constant error due to electronic noise and background
in the RGA, and a relative error due to effects that impact the
concentration in the measurement chamber.  The constant error, $\Delta
I_0 = 2.0 \times 10^{-12}$ A comes from the variation of the slope of
the undoped data sets, where two measurements with identical dopings
were possible.  The relative error, $\frac{\Delta I}{I} = 0.30$ comes
from the variation in the slopes of the doped data sets.  This
combination of errors yields a reduced $\chi^2$ of 0.96 and 1.04 for
the low and high doping fits respectively.

Figure \ref{subfig:fitall} shows the fit to all data, using the same
error treatment as in the fits to the subsets, giving a reduced
$\chi^2$ of 1.6.  
The fits to the different subsets of data yield
concentrations of 470 ppt, 420 ppt, and 100 ppt, from which we claim a
value on the intrinsic concentration of $c_0 = 330 \pm 200$ ppt.

Finally, to estimate the sensitivity of this measurement method, we
exploit the spread in the isotopic abundance in the undoped
measurements.  The isotopes with the lowest abundance are not
clearly visible above the background, but the isotopes whose current
is larger than the background can be used to determine the
sensitivity.  Since mass 80 is measured at the level of the
background, but not clearly above it, our sensitivity for detection is
likely above this level.  We thus use masses  82 and 83,
whose abundances are 11.6\% and 11.5\% respectively, which yeild a
sensitivity of 40 ppt.

\section{Conclusion}
\label{sec:Conclusion}

A new method for measuring trace impurities in xenon gas has been
described, where concentrations of krypton in xenon down to the sub
ppb level can be 
detected.  The key features of this measurement method are the
sensitivity increase obtained by temporarily reducing the pumping
speed at the measurement chamber by partially closing a custom-made
butterfly valve and
the minimal amount of xenon necessary for the measurement.  Using this
method, the krypton concentration in a sample of xenon from Air
Liquide was measured at $330 \pm 200$ ppt, and the sensitivity of
this measurement was estimated at 40 ppt.  An independent measurement
of the krypton concentration in the same Air Liquide gas was performed
by our colleagues from the Max Planck Institute for Nuclear Physics in
Heidelberg using gas chromotography yielding an
intrinsic concentration of $370 \pm 80$ ppt \cite{Sebastian}.  This
independent check matches our result well, thus confirming our
result. By further reducing the systematic uncertainties, we might be
able to achieve uncertainties close to the sensitivity limit for
future measurements.

Our measurement method will be used for several
applications.  First, it can be used as a screening device for
cryogenic distillation.  It can be used to measure the input krypton
concentration, which is expected to be in the ppb range.  It can also
be used to measure the krypton enriched xenon that comes from the
high concentration end of the distillation column, allowing for an
estimate of the purity of the clean gas by comparison with the
incoming gas.  Finally, it can be used as a leak detector for ultra
pure gas at the sub ppt level.  An air leak would result in a large
increase in the krypton concentration, which would be detectable with
the cold trap enhanced RGA measurement.  Thus one can monitor the
stability of an ultra clean system with a minimum expenditure of gas.  

In addition to measuring krypton concentrations, this setup can
measure any impurity whose boiling point is below that of nitrogen.
Although these measurements have not yet been explored in detail,
future work will address quantitative analyses of such impurities.

\section{Acknowledgement}

This work was supported by Deutsche Forschungsgemeinschaft.  We 
thank Alexander Fieguth, Michael Murra, and Martin Schlak for help
with construction and measurements.

\end{document}